\documentstyle[12pt,epsf]{article}

  \newcommand{\wdr}{(\sqrt{3} \times \sqrt{3})}
  \newcommand{\ztz}{p(2 \times 2)}
  \newcommand{\ave}[1]{\langle\,\mbox{$#1$}\,\rangle}
  \newcommand{\pssize}[1]{\setlength{\epsfxsize}{#1}}

\begin{document}
\title{\hfill \raisebox{20mm}[0mm][-20mm]{\normalsize ITP-UH 21/93}\\[-10mm]
       Eutectic point in a simple lattice gas model}

\author{M.~Sandhoff
        \thanks{Email: sandhoff@kastor.itp.uni-hannover.de},
        H.~Pfn\"ur$^{\dagger}$ and H.-U.~Everts\\
        \small Institut f\"ur Theoretische Physik, Universit\"at Hannover\\
        \small Appelstra\ss e 2, 30167 Hannover, Germany\\
        \small $^{\dagger}$Institut f\"ur Festk\"orperphysik,
               Universit\"at Hannover\\
        \small Appelstra\ss e 2, 30167 Hannover, Germany}

\date{\today}
\maketitle

\begin{abstract}
We investigate the phase diagram and the critical properties of the
adsorbate system sulphur/ruthenium(0001) in the coverage region
$\frac{1}{4} < \Theta < \frac{1}{3}$ by means of Monte-Carlo simulations
of a simple lattice gas model on a triangular lattice.  The model contains
only repulsive nearest and next-nearest neighbor interactions. Combining
results obtained by using both Glauber and Kawasaki kinetics in the
simulations we identify two tricritical points, three coexistence regions
and a eutectic point in the phase diagram of the system. Our results agree
with the findings of recent experimental work on S/Ru(0001). Furthermore
we are able to add certain details to the experimentally observed phase
diagram.
\end{abstract}

\noindent {PACS 0550 - Lattice theory and statistics; Ising problems}\\
\noindent {PACS 6470 - Phase equilibria, phase transitions,
                       and critical points}
\vspace{10mm}

The comparison of experimental phase diagrams and of experimental data for
the critical properties of adsorbed layers with the results of computer
simulations of appropriate lattice gas models can provide detailed
information about the effective lateral interactions between the adsorbed
particles.  However, the reliability of this method depends on the
validity of the lattice gas description for the system to be
considered. Therefore, this point requires careful experimental tests. The
inadequacy of the lattice gas description may be the reason for the
quantitative disagreement between the simulated \cite{Sel82} and the
experimental \cite{Imh82} phase diagrams of some systems.

For the system S/Ru(0001), careful investigations \cite{Den92,Heu93,Jue93}
have revealed that for coverages $0 < \Theta < \frac{1}{2}$ the S atoms
are attached to well defined surface sites. For $0 < \Theta< \frac{1}{3}$
only the the hcp site is occupied while for higher coverages, $\frac{1}{3}
< \Theta< \frac{1}{2}$, one finds adsorbate atoms on the fcc site as
well. Thus, for coverages up to $\Theta = \frac{1}{2}$ the description of
the system by lattice gas models appears to be justified. Hence, for this
system the comparison between Monte-Carlo (MC) simulations of such models
and the rather complex experimentally observed phase diagram shown in
Fig.~\ref{exphase} is well founded. Apart from the existence of five
different commensurate structures the most interesting features of the
phase diagram of S/Ru(0001) are the striped domain wall phase and the
coexistence region of two ordered structures which covers a large area in
the interval $\frac{1}{4} < \Theta < \frac{1}{3}$. To our knowledge, this
is the first adsorbate system in which such a prominent coexistence region
has been observed. In view of the complexity of this phase diagram it may
seem questionable whether the experimental observations can be reproduced
in detail by a reasonably simple lattice gas model.

In this letter we shall demonstrate by means of MC simulations that, in
fact, a surprisingly simple lattice gas model suffices for a quantitative
description of the experimental phase diagram for coverages $\frac{1}{4}
\le \Theta \le \frac{1}{3}$ for which the coexistence region (region C) is
observed. This area is bounded on the left by the homogeneous $\ztz$ phase
(B) and by the homogeneous $\wdr$ phase on the right (D)
(Fig.~\ref{exphase}). The ideal coverages (also referred to as critical
coverages) of the two phases are $\Theta = \frac{1}{4}$ and $\Theta =
\frac{1}{3}$ resp.. The coexistence region consists of homogeneous islands
with $\ztz$ and $\wdr$ order.

The lattice gas model with the smallest number of parameters which is able
to produce these two structures is
\begin{equation}
\label{hamilton}
  {\cal H_{{\rm LG}}} = \varphi_1\sum_{nn}  c_i c_j
                    + \varphi_2\sum_{nnn} c_i c_j\,.
\end{equation}
As usual, $c_i = 0,1$ denotes the occupation of the lattice site $i$, and
the sums run over the $N \times N$ sites of the triangular lattice for
nearest (nn) and next nearest neighbors (nnn).

Positive coupling constants $\varphi_i$ for the first- and second-nearest
neighbor interactions stabilize the $\ztz$ structure at $\Theta =
\frac{1}{4}$ (Refs.~\cite{Glo83,San93}), while the existence of the $\wdr
$ structure requires that $\varphi_2/\varphi_1 < \frac{1}{5}$, see
Ref.~\cite{Wal79}. The formation of $\ztz$ islands at low coverages,
Fig.~1, indicates that in a realistic lattice gas model for S/Ru(0001) a
weak attractive third neighbor interaction has to be included. As we have
checked this interaction is of little importance for coverages $\Theta >
\frac{1}{4}$. Therefore we ignore it in the present work. Similar models
have been considered in previous numerical studies of adsorbate systems on
square lattices \cite{Lan72} and of systems on triangular lattices
\cite{Mih77,Lan83,Glo83,Pie92,Roe83}. For the parameter ratio $\varphi_2 /
\varphi_1 = \frac{1}{10}$, a Monte-Carlo simulation of the model
(\ref{hamilton}) has been performed by Glosli and Plischke \cite{Glo83},
but a number of questions remained unresolved in their work.

Commonly, simulations have been performed with the use of Glauber kinetics
i.e. for constant chemical potential $\mu$ (Ref.~\cite{BinHer}). This
requires that a term $ \mu \sum c_i $ is added to the hamiltonian
(\ref{hamilton}) in order to control the coverage $\Theta = \frac{1}{N^2}
\sum_i c_i$. Practically, with this technique it turns out to be rather
difficult to tune to a coverage that deviates significantly from the ideal
coverages. In a Monte-Carlo run the system tends to jump between the
competing ideal structures, and very long runs are necessary to achieve
accurate statistics. Furthermore, this simulation technique does not
directly mimic the experimental situation. The experiments are done at
constant coverage and not at constant chemical potential. Thus it appears
appropriate to use Kawasaki kinetics \cite{Kaw72} which conserves the
coverage.

To obtain a detailed picture of the phase diagram in the intermediate
region between the ideal coverages $\Theta = \frac{1}{4}, \frac{1}{3}$ we
combine results from both types of simulation techniques.

As a measure for long range order of the $\ztz$ type and of the $\wdr$
type we use appropriate order parameters $\Psi_{\ztz}$ and $\Psi_{\wdr}$
for which explicit expressions can be found in Refs. \cite{San93} and
\cite{Due91}. In particular, these quantities allow the independent
measurement of the fraction of these structures which constitute the
coexistence region.

To characterize the nature of the phase transitions, we calculate the
susceptibilities $\chi_{\ztz},\, \chi_{\wdr}$ and the specific heat $c$,
\begin{eqnarray}
   \chi & = & \frac{N^2}{k_BT} \left(\ave{\Psi_N^2} - \ave{\Psi_N}^2\right)\\
   c    & = & \frac{N^2}{k_BT^2} \left(\ave{\varepsilon_N^2}
                                     - \ave{\varepsilon_N}^2\right)\,.
\end{eqnarray}
Here $\varepsilon$ is the energy per site and $\ave{\ldots }$ denotes the
average over Monte-Carlo sweeps. The critical divergences of $c$ and
$\chi$ in the infinite system appear as maxima for finite $N$.  From the
size dependence of these maxima the critical exponents $\alpha, \gamma,
\nu$ and the transition temperature $T_c$ can be determined
\cite{Nig76,Bar83}.

In simulations at constant chemical potential the coverage $\Theta$
fluctuates as a function of the temperature
\begin{equation}
   \Delta\Theta := \frac{N^2}{k_BT}
                  \left(\ave{\Theta_N^2} - \ave{\Theta_N}^2\right)\,.
\end{equation}
The quantity $\Delta\Theta$ does not necessarily couple to the order
parameter. However, it is clearly of interest for first order phase
transitions: While the chemical potential and the temperature have to be
equal in both phases the coverage jumps discontinuously at the
transition.

In our simulations we work with triangular lattices with periodic boundary
conditions. The lattice sizes vary between $12 \times 12$ and $60 \times
60$ sites.  Because of the different unit cells of the $\ztz$ and and of
the $\wdr$ structures we have to change the linear dimension $N$ in steps
of $12$ to avoid frustration by boundary effects.  Each data point is
obtained from $600\,000$ to $1\,200\,000$ Monte-Carlo sweeps.

The ratio of the order-disorder transition temperatures at the ideal
coverages $\Theta= \frac{1}{4}, \frac{1}{3}$ measured in the experiment
\cite{Den92} is used to fix the coupling constant $\varphi_2$ in units of
$\varphi_1$. As usual we set $\varphi_1 = 1$ and measure all energies in
units of this nn coupling. This allows the direct comparison of results
from both types of simulations. Agreement with the experiment is
established with $\varphi_2 = 0.123$.

At the coverages $\Theta = \frac{1}{4} , \frac{1}{3}$ both kinetics can be
used for the determination of critical properties because the chemical
potential is practically constant across the transition temperature. By
comparison of the critical exponents with the results obtained from the
experiment and with the values predicted by the Landau rules
\cite{LanLif,Sch81} we are able to check our routines. Our finite size
analysis reveals that Glauber kinetics and Kawasaki kinetics lead to
identical results for both coverages: The maximum of the specific heat at
$\Theta = \frac{1}{4} (\frac{1}{3})$ has a size dependence $c_N^{max} \sim
N^{\frac{\alpha}{\nu}}$ with $\frac{\alpha}{\nu} = 1.10\pm0.1
(0.50\pm0.06)$, see Fig.~\ref{c025}, while the susceptibility $(
\chi_N^{max} \sim N^{\frac{\gamma}{\nu}} )$ yields $\frac{\gamma}{\nu} =
1.95\pm0.2 (1.65\pm0.2)$.  These exponents are comparable with the exact
values of the $q=4$ ($q=3$) Potts model $\frac{\alpha}{\nu} = 1 (0.4),
\frac{\gamma}{\nu} = 1.75 (1.7\bar{3})$ (Ref.~\cite{Wu82}) and with the
experimental values. Previous simulations of lattice gas models with
$\ztz$ and $\wdr$ symmetries have yielded similar differences between the
lattice gas exponents and the Potts exponents
\cite{Mih77,Lan83,Glo83,Pie92,Roe83}.

As we have discussed above, Kawasaki kinetics is the appropriate tool to
analyze the ordered structures in the intermediate coverage region,
$\frac{1}{4} < \Theta < \frac{1}{3}$, and the boundaries between these
structures. For sufficiently low temperatures we find that particles added
to the $\ztz$ structure do not appear as point defects but as islands of
$\wdr$ order.  Due to the interfacial tension this {\em minority} phase
forms hexagons in a sea of the ideal $\ztz$ phase ({\em majority}
phase). At $\Theta \simeq 0.292 \simeq \frac{1}{2}\left(\frac{1}{4} +
\frac{1}{3}\right)$ both domains occupy half the available area.  Above
$\Theta = 0.292$ the roles are interchanged, i.e. the denser $\wdr$
structure becomes the majority phase which surrounds islands of $\ztz$
phase. The coexistence line of the region III, Fig.~(\ref{myphase}), in
which the $\ztz$ and the $\wdr$ structure coexist, is determined by the
disappearance of the order parameter measuring the minority domains.

Clearly the boundaries of region III must end at the ideal coverages for
$T=0$. We are unable to confirm this for $\Theta < 0.27$ and $\Theta >
0.31$ because the area covered by the minority phase is too small.

To avoid Fisher renormalization \cite{Fis68} we investigate the
order-disorder transition of the majority phases, $\ztz$ and $\wdr$ resp.,
with Glauber kinetics.  Paths of constant $\mu$ are shown in
Fig.~\ref{myphase}. As $\Theta$ approaches $\Theta = 0.292$ the
N-dependence of the finite-size peak in the specific heat $c \sim
N^{\frac{\alpha}{\nu}}$ deviates significantly from the Potts values
$\frac{\alpha}{\nu} = 1, \frac{2}{5}$, resp.. For $\mu=-0.50$, label a,
which corresponds to the ideal coverage $\Theta = \frac{1}{4}$, the
exponent is $1.10$, see above. For $\mu=-0.58$ we find $1.33$ (label b)
and on the path $\mu=-0.68$ we obtained $1.67$ (label c).  As $\Theta$
approaches $\Theta = 0.292$ from higher coverages $\frac{\alpha}{\nu}$
increases from $0.5$ for $\mu = -0.98$, to $1.1$ for $\mu = -0.74$ (paths
d - f). This behavior of $\frac{\alpha}{\nu}$ is indicative of the
existence of tricritical points on the order to disorder transition
lines. In the interval between these tricritical points $\Theta_{\rm
tr}^{(1)} < \Theta < \Theta_{\rm tr}^{(2)}$, the transition is a
fluctuation induced first order transition. While the coverage remains
close to the ideal coverages $\Theta = \frac{1}{4}, \frac{1}{3}$ resp. on
lines of constant chemical potential which cross the phase boundary
outside the interval $\Theta_{\rm tr}^{(1)} < \Theta < \Theta_{\rm
tr}^{(2)}$, it changes rapidly on lines $\mu = {\rm const.}$ that cross
the phase boundary within this interval.  On these last lines the coverage
develops large fluctuations.  This is seen in Figs.~(\ref{dt1}) and
(\ref{dt2}) where we plot $\Delta\Theta$ versus $kT$. Evidently, for
$\Theta_{\rm tr}^{(1)} < \Theta < \Theta_{\rm tr}^{(2)}$ the peaks in
$\Delta \Theta$ scale with a power of the system size. By contrast,
$\Delta \Theta$ is structureless in the vicinity of the ideal coverages,
where the transition is of second order.  The insets in Fig.~\ref{dt1} and
Fig.~\ref{dt2} display the jump in coverage on lines of constant $\mu$ and
give an estimate of the width of regions I, II in which the $\ztz$ phase
and the $\wdr$ phase coexist with the lattice gas phase. From this point
of view the maxima of the specific heat and of the susceptibilities
obtained from simulations are to be considered as averages over these
coexistence regions. The determination of the tricritical points from
finite size results is difficult since the exponents $\frac{\alpha}{\nu},
\frac{\gamma}{\nu}$ increase smoothly when $\Theta$ is varied between the
ideal coverages and $\Theta = 0.292$.  In general, the tricritical points
are the lower (upper) end points of the coexistence regions I and II. We
determine the boundaries of these coexistence regions by extrapolating the
points of maximum curvature of the graphs $\Theta(T)$ shown in
Fig.~\ref{dt1} and Fig.~\ref{dt2}.  This provides the estimate for the
locations of the tricritical points in Fig.~\ref{myphase}.

The two upper boundaries of the coexistence regions I,II must meet the
boundary of the coexistence region III in one point, i.e. at $\Theta =
0.292$. Otherwise the region III would have a coexistence line with the
lattice gas phase at which three phases coexist. This would contradict
Gibbs' phase rule \cite{LanLif} which, for a two component system with
three phases, allows only for a triple point.  This point is well known
for binary alloys in 3d and is called {\em eutectic point}.  We check the
first order nature of the phase transition at the eutectic point with
simulations for constant $\Theta$.  For symmetry reasons the chemical
potential is constant on the line of constant coverage $\Theta =
0.292$. Thus, in contrast to the general case $\Theta \ne 0.292$,
calculations of critical exponents on this line are not affected by Fisher
renormalization \cite{Fis68}. At the eutectic point our simulations yield
the following size dependence of the specific heat and the
susceptibilities: $c \sim N^{1.8}$, $\chi_{\ztz} \sim
N^{1.32}$,$\chi_{\wdr} \sim N^{1.52}$. For first order transitions these
thermodynamic quantities are predicted to diverge as $N^2$ with the system
size \cite{Cha86}. The reason for the discrepancy between this theoretical
prediction and our numerical values for the exponents of the
susceptibilities is visible in Fig.~\ref{c-chi029}. While the positions of
the peaks of the specific heat, $c_N^{(\max)}$, increase linearly with
$\ln(N)$, the peaks of the susceptibilities, $\chi_N^{(\max)}$, curve
upwards as function of $\ln(N)$.  This means that with our system sizes we
have not yet reached the scaling regime. In fact, the data for
$\chi_{\ztz}$ and $\chi_{\wdr}$ are based on effective system sizes
$N_{\rm eff} < \frac{N}{2}$ because for $\Theta = 0.292$ the $\ztz$ and
the $\wdr$ structures cover less than half the lattice.

In summary, we have shown that the simple lattice gas model
(\ref{hamilton}) yields a surprisingly rich phase diagram in two
dimensions which closely resembles phase diagrams commonly observed in
three dimensional substitutional binary alloys. It accounts for all
details of the experimental phase diagram in the coverage region
$\frac{1}{4} \le \Theta \le \frac{1}{3}$. This includes the in particular
coexistence regions I and II.  These regions and their structure could not
be determined with certainty in the experiment although their existence
had been inferred from the structure of the experimental phase diagram in
the vicinity of $\Theta = 0.292$.  Finally, we note that the topology of
the phase diagram found in this study can be expected to occur more
generally for systems with threefold symmetry provided the lateral
interactions decrease with distance comparably fast as in our model.

\section*{Acknowledgements}
This work has been supported by the Deutsche Forschungsgemeinschaft under
grant no. Pf~238/2.

\newpage
{\sc

}

\newpage
\section*{Figure Captions}

\begin{figure}[h]
\caption[]{Experimental phase diagram of the system S/Ru(0001) \cite{Sok92}.
A: $\ztz$ islands $+$ lattice gas.
E': $\wdr$ $+$ line defects.
Regions A, E and E' are not considered in the simulation.}
\label{exphase}
\end{figure}

\begin{figure}[h]
\caption[]{Specific heat c for fixed coverage $\Theta = \frac{1}{4}$; the
curves correspond to the system sizes $24 \times 24$, $36 \times 36$, $48
\times 48$ and $60 \times 60$. The inset displays the linear fit through
the maxima of $c_N$.}
\label{c025}
\end{figure}

\begin{figure}[h]
\caption[]{Calculated phase diagram. Solid lines: boundaries of different
regions.  Region I \{II\}: coexistence of the disordered phase with the
$\ztz$ \{$\wdr$\}\ phase. Region III: coexistence of $\ztz$ and
$\wdr$. Tricritical points: $\Theta_{tr}^{(1)} = 0.256,\quad kT_{tr}^{(1)}
= 0.173$; $\Theta_{tr}^{(2)} = 0.324,\quad kT_{tr}^{(2)} = 0.170$ marked
by $\bullet$.  Dotted lines: maxima of specific heat $c$ for constant
$\Theta$. Dash dotted lines: paths of $\mu = {\rm const.}$.}
\label{myphase}
\end{figure}

\begin{figure}[h]
\caption[]{Coverage fluctuations $\Delta\Theta$ at first order phase
transitions for three different values of $\mu$ corresponds to the labels
(a - c) in Fig.~(\ref{myphase}). Curves are shown for the same system
sizes as in Fig.~(\ref{c025}).}
\label{dt1}
\end{figure}

\begin{figure}[h]
\caption[]{Same as in Fig.~\ref{dt1} but for $\Theta > 0.292$.}
\label{dt2}
\end{figure}

\begin{figure}[h]
\caption[]{Specific heat $c$ (a); susceptibilities $\chi_{\ztz},
\chi_{\wdr}$ (b) at the eutectic point. Insets: positions of the maxima of
$c,\chi_{\ztz}$ and $\chi_{\wdr}$ as functions of the lattice size
$N$. System sizes are in Fig.~(\ref{myphase})}
\label{c-chi029}
\end{figure}
\clearpage
\newpage

\section*{Figures}
\parindent0mm \newlength{\baselength} \baselength170mm
\centerline{ \pssize{\baselength} \epsfbox{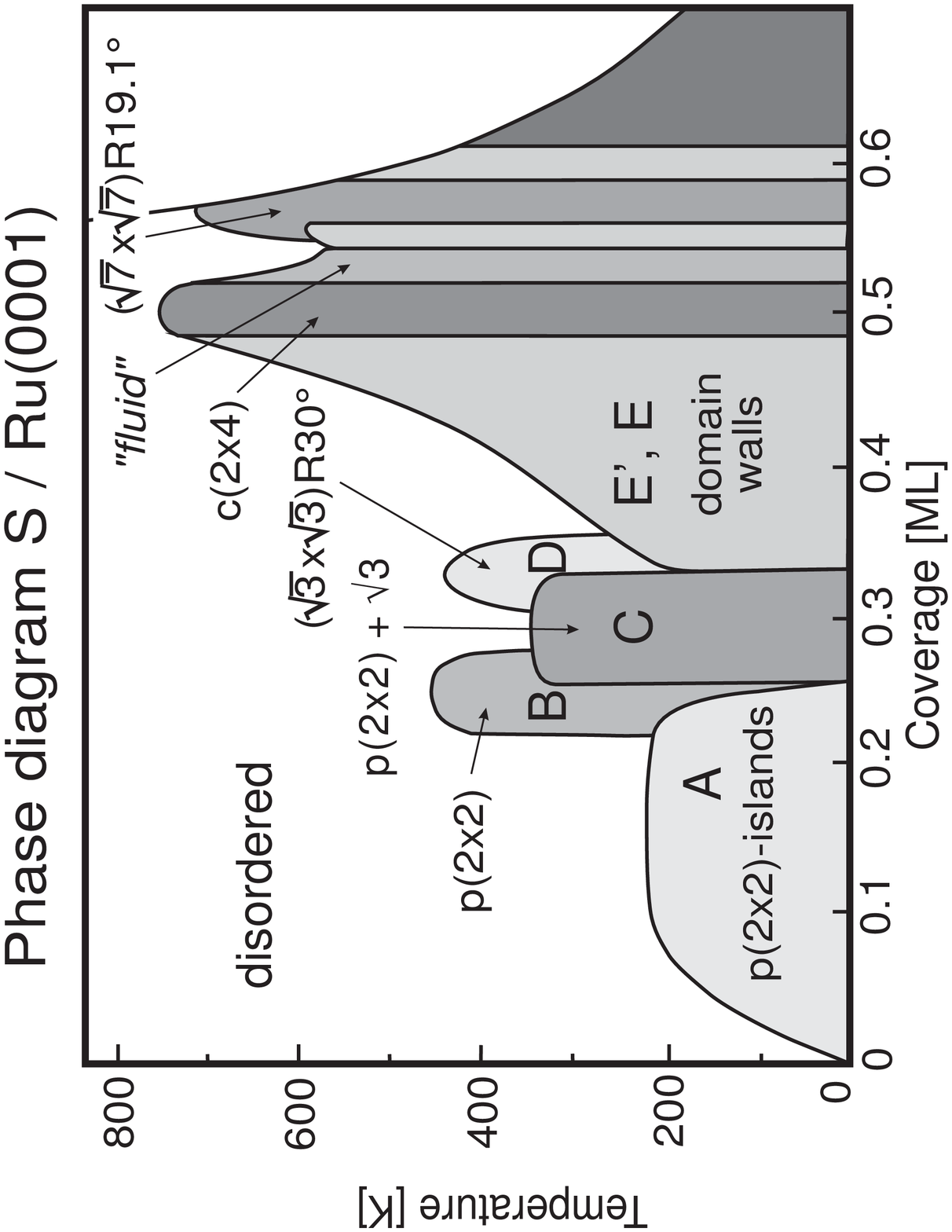} }
\centerline{ \pssize{\baselength} \epsfbox{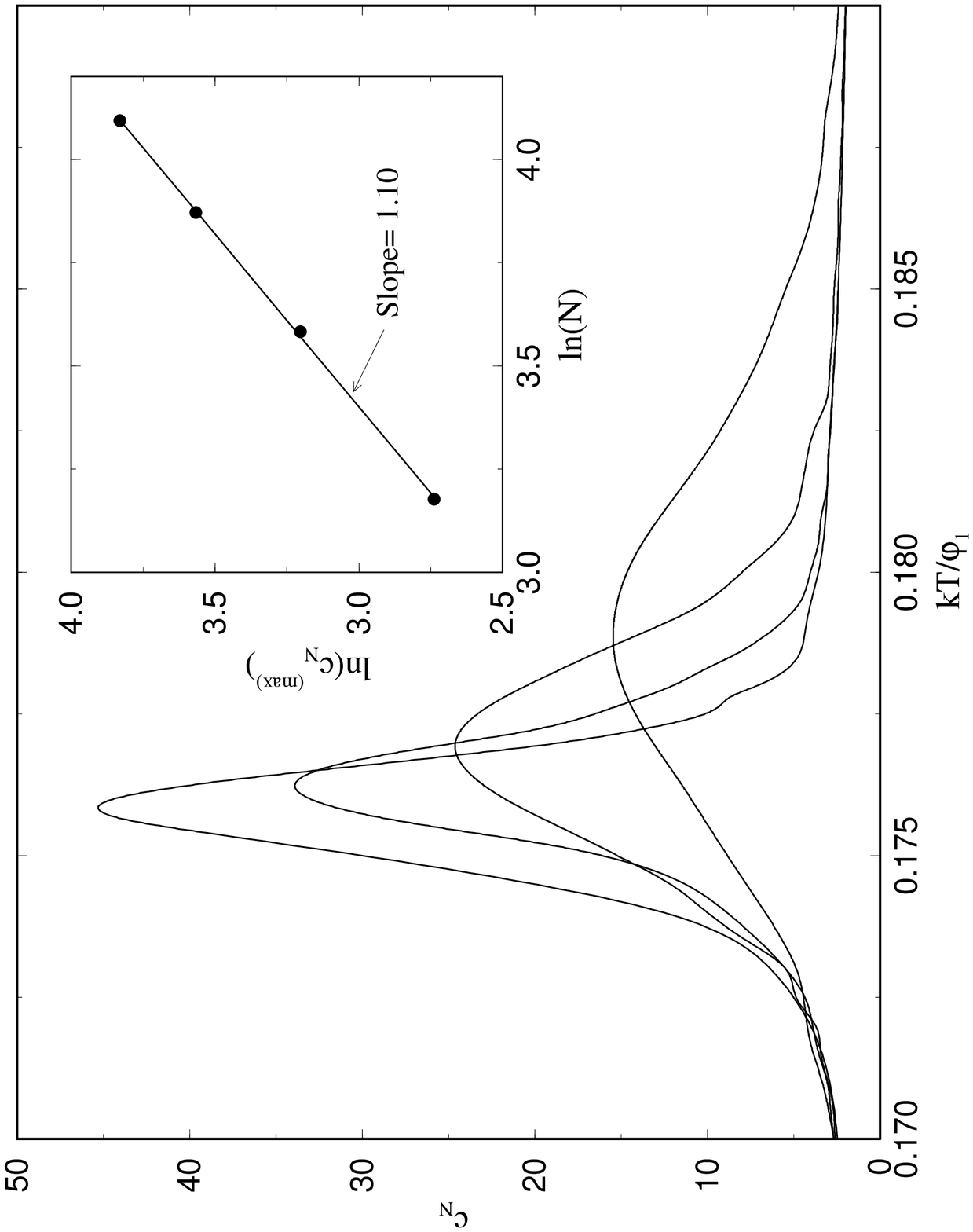} }
\centerline{ \pssize{\baselength} \epsfbox{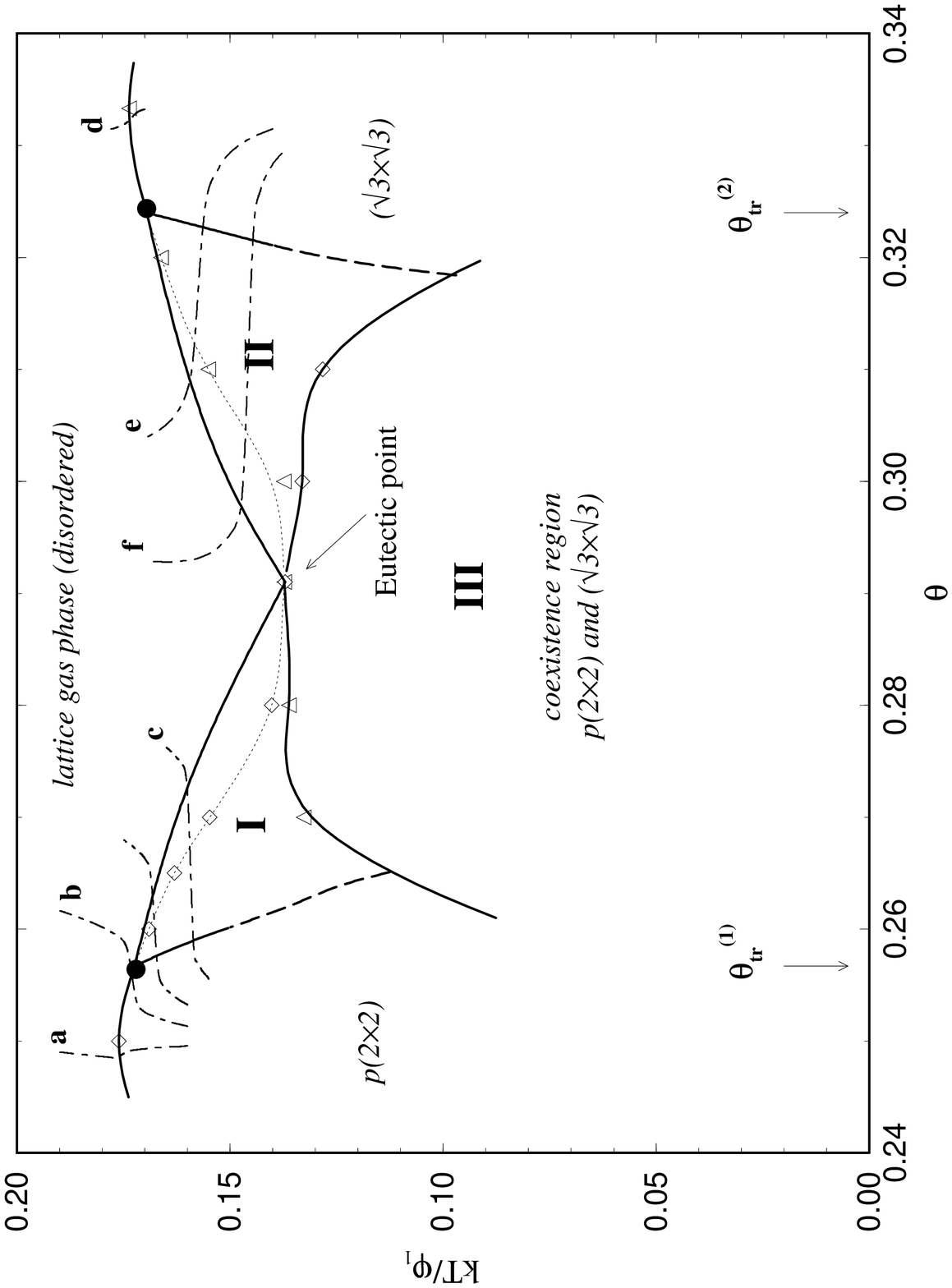} }
\centerline{ \pssize{\baselength} \epsfbox{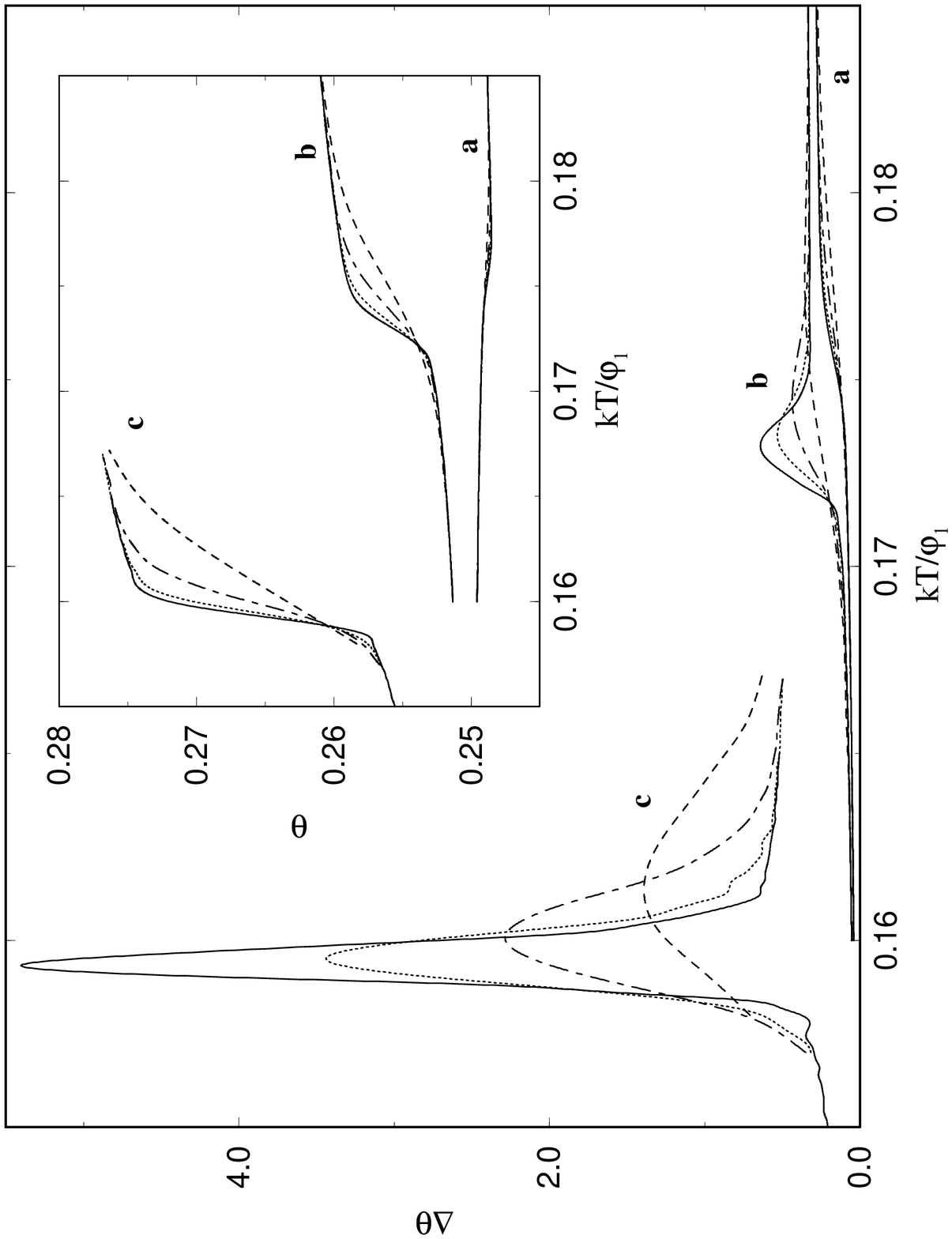} }
\centerline{ \pssize{\baselength} \epsfbox{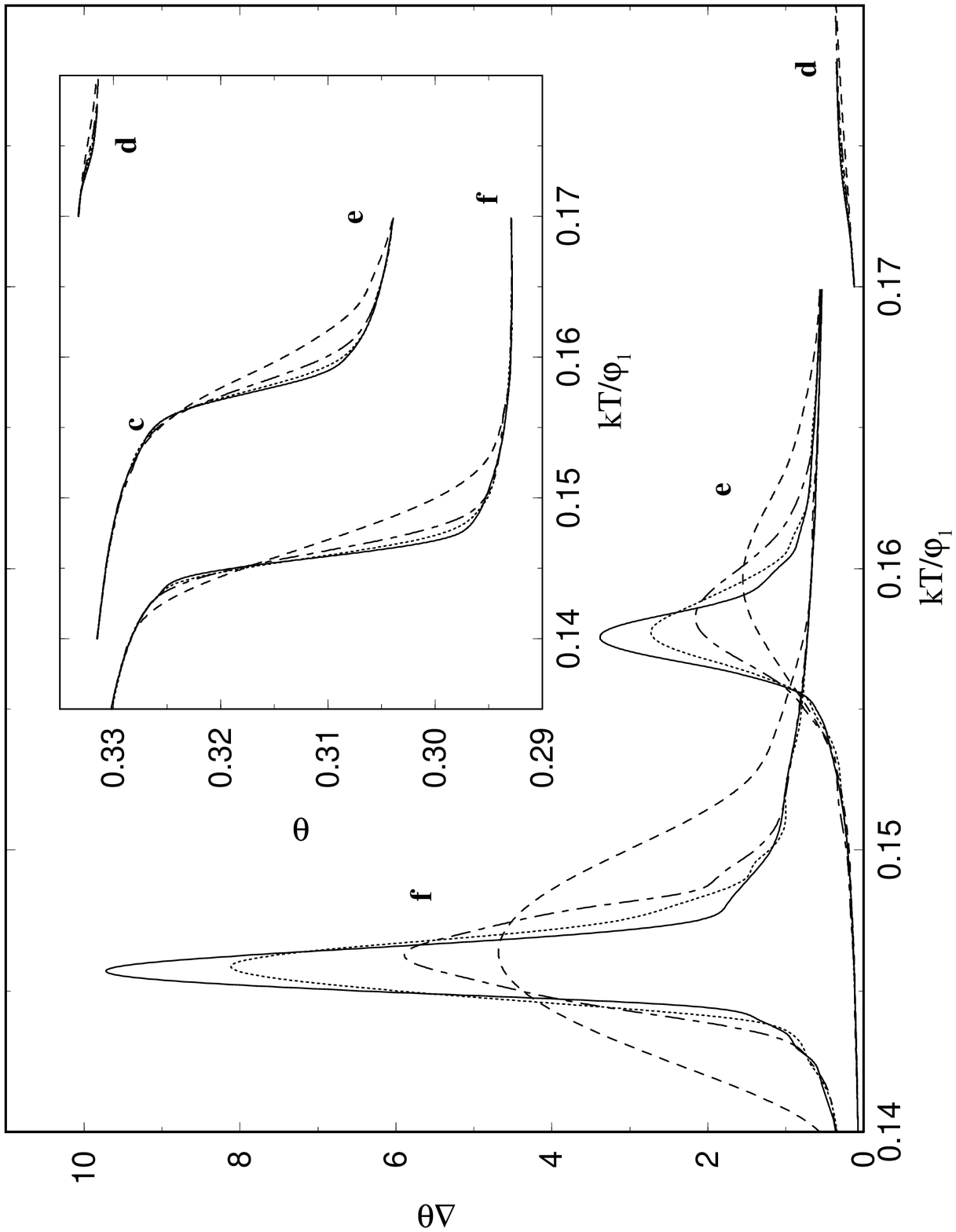} }
\centerline{ \pssize{\baselength} \epsfbox{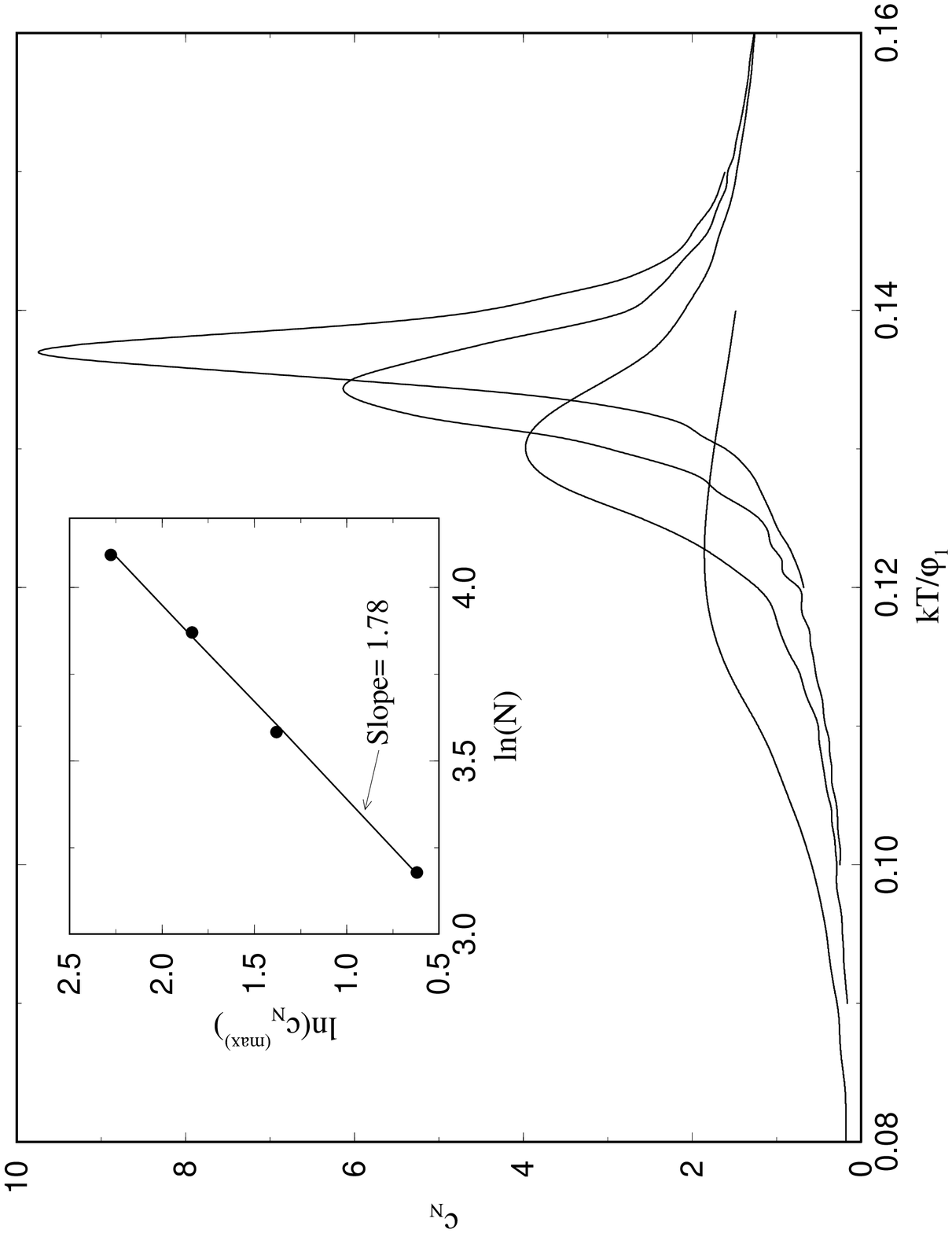} }
\centerline{ \pssize{\baselength} \epsfbox{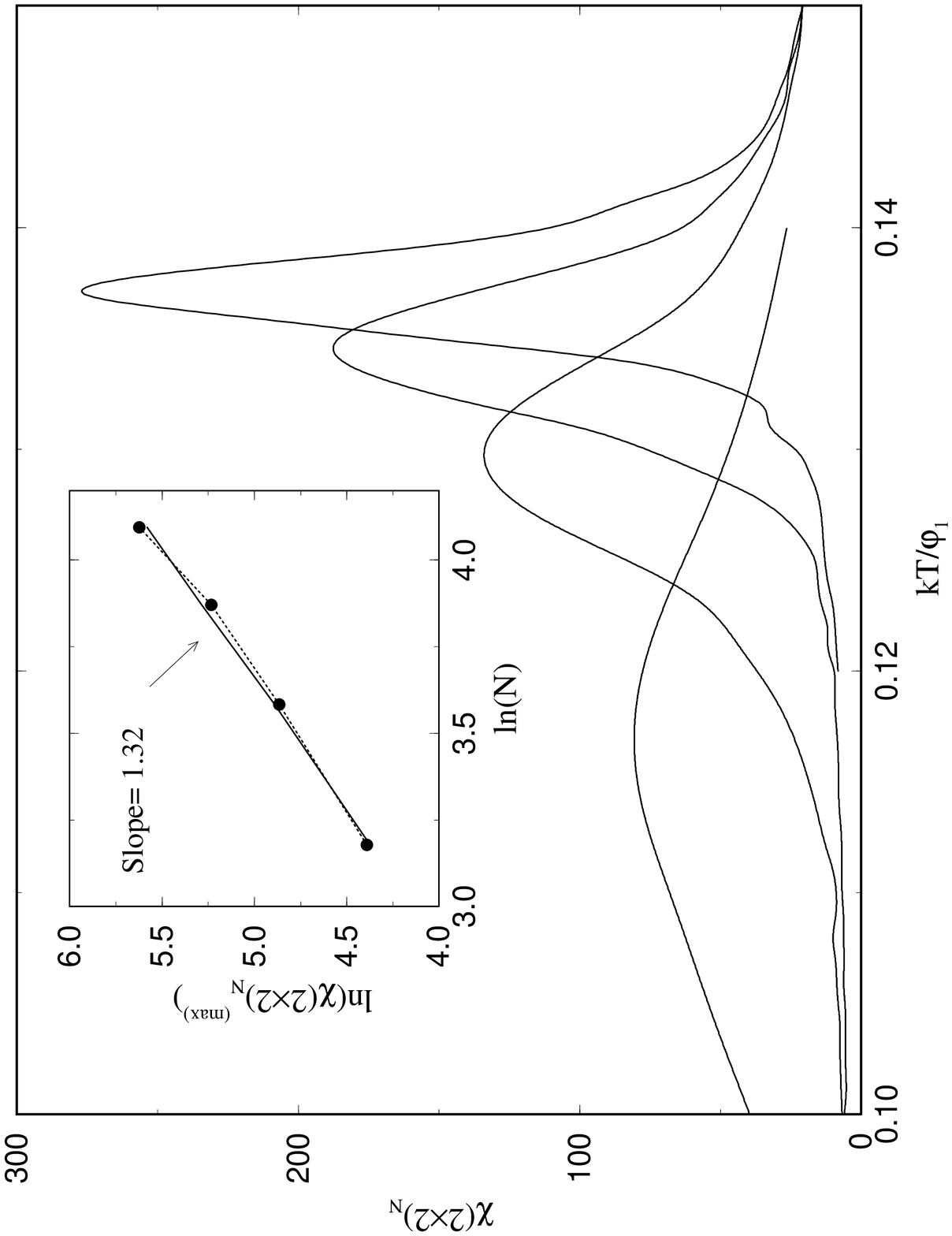}}
\centerline{ \pssize{\baselength} \epsfbox{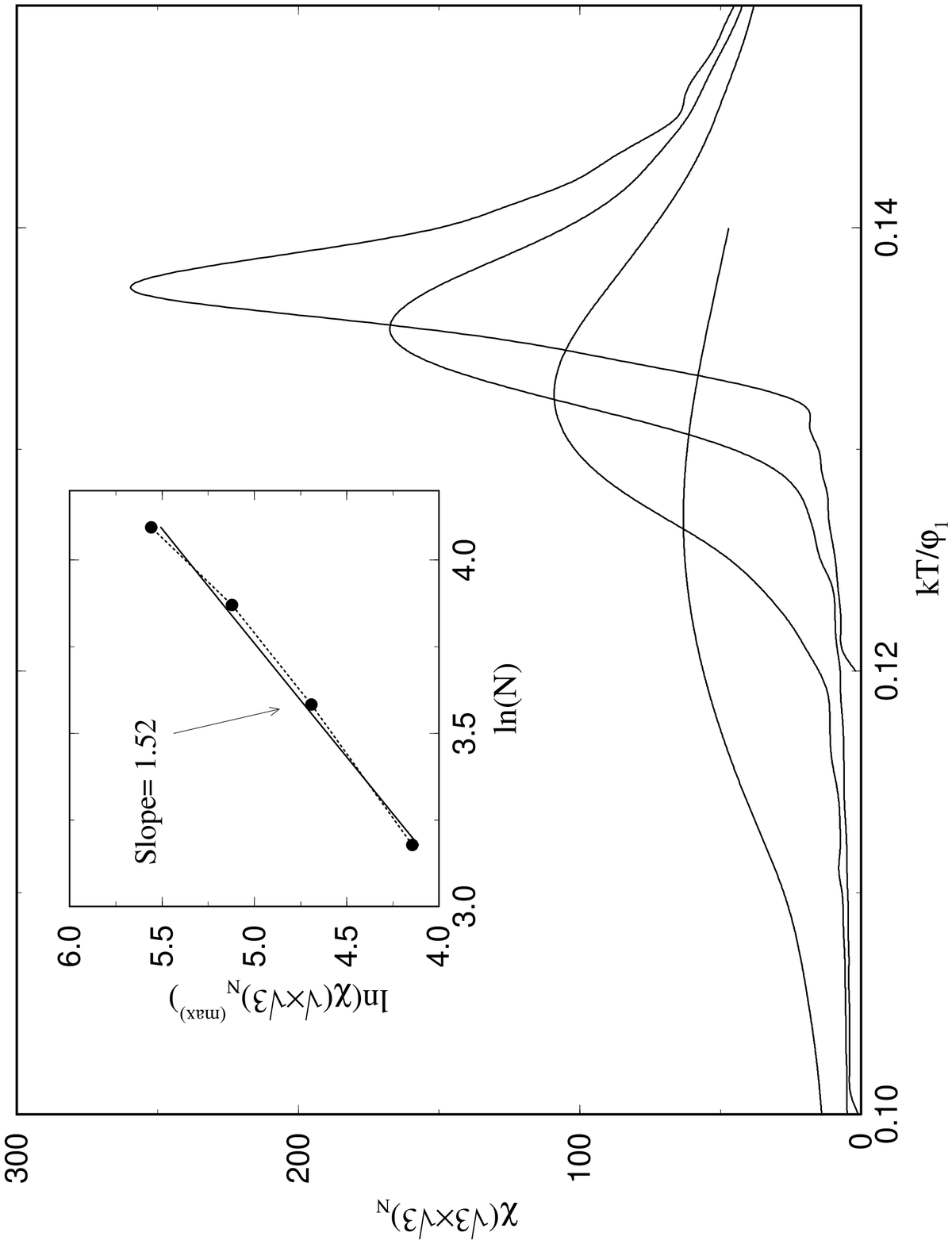} }
\end{document}